\documentclass[journal,9pt,twocolumn]{IEEEtran}

\usepackage{cite}
\usepackage{amsmath,amssymb,amsfonts}
\usepackage{graphicx}
\usepackage{textcomp}
\usepackage{xcolor}
\usepackage{booktabs}
\usepackage{microtype}
\usepackage{hyperref}
\usepackage{placeins}

\begin{document}

\title{IMU-Aided Correction of Orientation-Induced Ranging Error
in Bluetooth Channel Sounding on Commercial Hardware}

\author{Mihir~Bapat$^{1}$ and Santosh~Nagaraj$^{2}$%
\thanks{This work has been submitted to the IEEE for possible
publication. Copyright may be transferred without notice, after which
this version may no longer be accessible.}\\
\small $^{1}$Del Norte High School, San Diego, CA 92127 USA\\
\small $^{2}$Department of Electrical and Computer Engineering,
San Diego State University, San Diego, CA 92182 USA\\
\small Corresponding author: M. Bapat (e-mail: mihirnbapat@gmail.com).}

\maketitle

\begin{abstract}
Bluetooth Low Energy Channel Sounding (BLE CS), standardized in Bluetooth Core Specification 6.0 (September 2024), enables distance estimation via phase-based ranging (PBR) and round-trip time (RTT). Although prior work has studied CS accuracy in configurations with a fixed orientation, no published work has studied how device orientation affects ranging error on commercial hardware. We present the first study of orientation-induced CS ranging error and a Machine Learning correction using IMU features. This study used the EFR32xG24 Channel Sounding Development Kit, the only commercial CS platform with an integrated six-axis Inertial Measurement Unit (IMU). Our results show that device orientation has a substantial effect on CS ranging accuracy; we found that a Random Forest model trained on the IMU derived orientation achieved a 74.6\% Mean Absolute Error (MAE) reduction under a Leave One Orientation Out Evaluation, demonstrating that IMU readings have potential to improve ranging accuracy.
\end{abstract}


\section{Introduction}

\IEEEPARstart{B}{luetooth} Channel Sounding (CS) is a significant
improvement in short-range wireless distance measurement, as it
estimates sub-meter distances with standard Bluetooth Low Energy (BLE)
radios without the need of complex antenna arrays required by
angle-of-arrival or angle-of-departure methods~\cite{r1}. Standardized
in September 2024 as part of Bluetooth Core Specification 6.0~\cite{r2},
CS has been of interest in applications such as keyless vehicle entry,
digital access keys, asset tracking, and proximity services: all
contexts in which devices are carried in varying and uncontrolled
orientations~\cite{r3,r4}.

BLE CS has two defined ranging methods. Phase-Based Ranging (PBR)
estimates distance from the accumulated phase difference of signals
exchanged across up to 72 channels in the 2.4~GHz band, while
Round-Trip Time (RTT) estimates distance from the round-trip travel
time of the signal~\cite{r3,r5}. Both are sensitive to multipath
propagation, which is determined in part by the radiation pattern of
the device antenna. This pattern changes significantly when the device
is rotated or tilted~\cite{r6}. Silicon Labs, whose hardware is used
in our study, acknowledges in their antenna design guidelines that
device orientation has a significant impact on CS ranging accuracy,
and that single-antenna configurations can produce errors of several
meters depending on how the device is oriented~\cite{r7}.

Despite the impact orientation can have on accuracy, to the best of our
knowledge, all published CS accuracy studies keep device orientation
constant during measurement. Wieme et~al.~\cite{r8}, the most
comprehensive hardware evaluation of BLE CS to date, identified the
onboard Inertial Measurement Unit (IMU) as a potential method for
compensating for orientation-based ranging errors and explicitly left
this as an open direction for future work. To the best of our
knowledge, no subsequent study has addressed this gap.

\begin{figure*}[t]
\centering
\includegraphics[width=\textwidth]{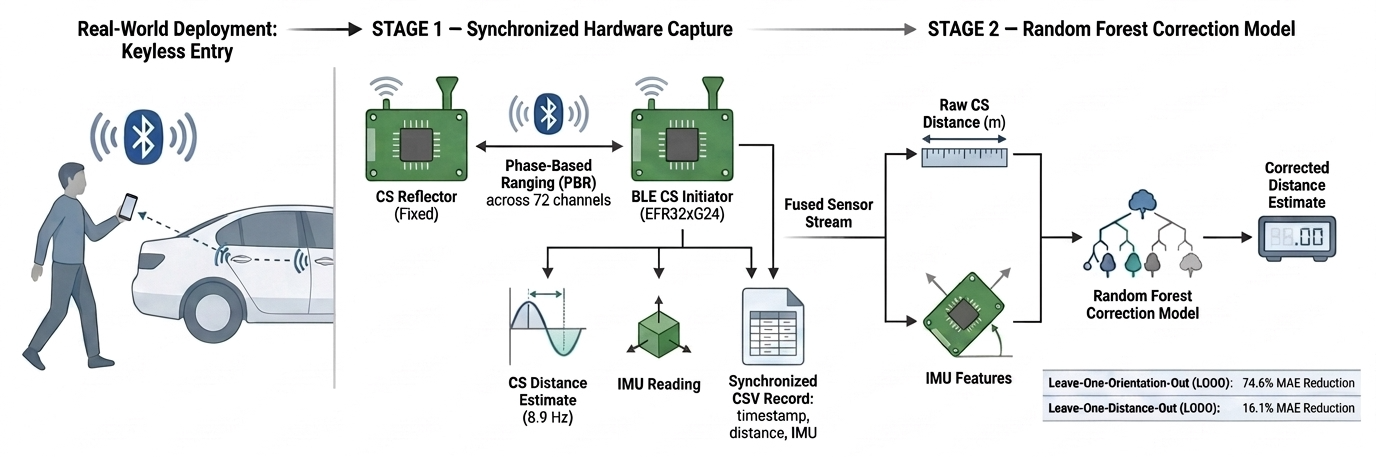}
\caption{Overview of the measurement and correction pipeline. The
EFR32xG24 initiator board synchronously captures CS distance estimates
and six-axis IMU readings at 8.9~Hz. These are combined with
ground-truth distance labels to characterize orientation-induced
ranging error and train a Random Forest correction model evaluated via
Leave-One-Orientation-Out cross-validation.}
\label{fig:pipeline}
\end{figure*}

This paper makes four contributions: (1) the first characterization of
orientation-induced CS ranging error across nine orientations and nine
distances; (2) the identification of a consistent asymmetry in ranging
error between roll and pitch orientations; (3) a statistically
significant correlation between IMU tilt and ranging error ($r=0.156$,
$p<0.001$); and (4) a Random Forest correction model showing 74.6\%
Mean Absolute Error (MAE) reduction after Leave-One-Orientation-Out
(LOOO) evaluation across all nine orientations, demonstrating that IMU
features generalize to unseen orientations. The system pipeline is
illustrated in Fig.~\ref{fig:pipeline}.

The remainder of the paper is organized into 5 sections. Section~II
covers related work on CS ranging, Machine Learning based correction
approaches, and antenna orientation effects. Section~III describes the
methodology, particularly the hardware setup, data collection procedure,
and feature computation. Section~IV provides the results of this study,
Section~V analyzes these results by discussing our findings, and
Section~VI concludes the paper with future studies.

\section{Background and Related Work}

There have been attempts at phase-based ranging for BLE before
standardization by Zand et~al.~\cite{r9}. Woolley~\cite{r3} defines
PBR and RTT modes in Bluetooth~6.0. Gunia and Ellinger~\cite{r10}
compare CS against UWB and FMCW radar. Recent work has evaluated CS
for vehicle access~\cite{r4} and in challenging indoor
environments~\cite{r11}.

The use of Machine Learning, specifically parametric neural networks,
has been explored in correcting CS measurements~\cite{r12}. MVDR-based
pipelines have also been studied~\cite{r13}, with successful RMSE
reductions of up to 0.4~m. It has also been shown that combining neural
networks across several CS measurements improves accuracy~\cite{r14}.
Similar Machine Learning approaches for error correction in BLE
multipath have been demonstrated~\cite{r16,r17}, and location-aware
error correction has reduced UWB P90 ranging error by 58\% to 15~cm on
unseen trajectories~\cite{r15}. However, none of these approaches solve
orientation-induced ranging error, which is the novelty of this study.

There have been some studies showing the effect of antenna orientation.
Dashti et~al.~\cite{r18} showed that directional antennas reduce UWB
ranging error. Pasku et~al.~\cite{r19} showed that rotation about
specific axes produces errors in RF space-diversity systems.
Bou-El-Harmel et~al.~\cite{r20} showed that antenna orientation,
including changes in polarization, changes indoor multipath
characteristics in sensor networks. No peer-reviewed study isolates the
effects of device orientation on CS measurements.

\section{Methodology}

\subsection{Hardware and Firmware}

In this study, two commercially available EFR32xG24 Channel Sounding
Development Kits (BRD2606A) were used. One board was flashed with the
standard Silicon Labs Reflector image and connected to a dormant
laptop, while the other was flashed with a customized initiator image
that reads and outputs the IMU measurements. The initiator was
connected to a host laptop, which ran the Channel Sounding measurements
in PBR mode with dual-antenna polarization diversity configured on both
boards. The RTL library produced filtered measurements at approximately
8.9~Hz.

The default initiator firmware was modified to read the onboard
ICM-40627 measurements immediately after each CS measurement in the
main application loop. Blocking the BLE event callback with SPI reads
caused buffer exhaustion on the reflector board; moving the IMU read to
the main loop avoids this. For every entry recorded, the following
features were captured: timestamp~(ms), CS distance~(m), $a_x, a_y,
a_z$ (raw ADC, $\pm$16g, 2048~LSB/g), $g_x, g_y, g_z$ (raw ADC,
$\pm$2000$^\circ$/s, 16.4~LSB/$^\circ$/s).

\subsection{Data Collection}

Data was collected in an indoor, straight-line corridor with the
reflector fixed at one end and the initiator placed at different
distances. Nine distances were tested: 1~ft (0.30~m), 2~ft (0.61~m),
3~ft (0.91~m), 5~ft (1.52~m), 8~ft (2.44~m), 12~ft (3.66~m),
18~ft (5.49~m), 25~ft (7.62~m), and 30~ft (9.14~m). For each
distance, nine orientations were tested: \textit{flat} (face-up,
reference), \textit{roll left 30}, \textit{roll right 30},
\textit{roll left 90} (vertical on long edge), \textit{pitch toward
30}, \textit{pitch away 30}, \textit{pitch toward 60}, \textit{face
down}, and \textit{upright} (vertical on short edge). Each
configuration was recorded for 60 seconds at 8.9~Hz, yielding between
520 and 1,185 samples. In total, 44,576 measurements were recorded
across 81 sessions. Throughout all sessions, the reflector was kept in
a fixed flat orientation. The tile angle was calculated from the raw accelerometer readings using standard triaxial accelerometer tilt formulas~\cite{r25}, while gyroscope magnitude and ranging error were found directly from the raw gyroscope and CS distance measurements.

\section{Results}

\subsection{Overview}

The overall MAE across all measurements was 164.17~cm and the RMSE was
207.24~cm, primarily due to increased error at longer distances. At
shorter distances between 1 and 3~ft, CS measurements are relatively
accurate (MAE 19.70 to 39.15~cm). However, beyond 5~ft where indoor
multipath becomes substantially worse, accuracy decreases severely,
with MAE between 195 and 266~cm. This is consistent with prior indoor
CS studies~\cite{r8,r9}.

\subsection{Orientation Effect at Short Range}

Across all orientation and distance combinations at short range, the
lowest MAE was 9.29~cm at 2~ft pitching 30 degrees toward the
reflector, which closely approaches the 10~cm ranging target stated by
Hillyard et~al.~\cite{r5}. The highest short-range MAE was 57.57~cm,
rolling left 90 degrees at 3~ft. At a fixed distance of 2~ft,
different orientations produced a nearly 5-fold difference in MAE,
from 9.3~cm (pitch toward 30) to 44.6~cm (roll left 90), demonstrating
the impact orientation alone can have on ranging error.
Table~\ref{tab:mae} shows the key orientation statistics at 1 to 3~ft.

\begin{table}[h!]
\caption{Key Per-Orientation MAE at Short Range. \textbf{Bold} = best
per distance, \textit{italic} = worst.}
\label{tab:mae}
\centering
\small
\setlength{\tabcolsep}{4pt}
\begin{tabular}{llrr}
\toprule
Dist. & Orientation & Mean err. (cm) & MAE (cm) \\
\midrule
1~ft & flat            & $-19.2$ & 19.2 \\
1~ft & roll left 30    & $+1.5$  & \textbf{13.4} \\
1~ft & pitch away 30   & $-14.3$ & \textit{23.6} \\
\midrule
2~ft & flat            & $-25.3$ & 36.3 \\
2~ft & pitch toward 30 & $-4.7$  & \textbf{9.3} \\
2~ft & roll left 90    & $-12.4$ & \textit{44.6} \\
\midrule
3~ft & flat            & $+41.7$ & 44.9 \\
3~ft & pitch away 30   & --      & \textbf{16.3} \\
3~ft & roll left 90    & $-51.2$ & \textit{54.6} \\
\bottomrule
\end{tabular}
\end{table}

\FloatBarrier

The MAE is consistent across all orientations at 1~ft, with roll left 30
being the only orientation that produced a positive mean error. At
2~ft, pitching 30 degrees toward the reflector substantially
outperformed other orientations, whereas rolling left 90 degrees showed
high mean error and high variance. This high error in rolling left 90
degrees continued at 3~ft, suggesting a consistent
increase in error at this orientation.

Rolling left 90 degrees at 18 and 25~ft showed the highest MAE
across the set of orientation and distance combinations.

\subsection{Roll vs. Pitch Asymmetry}

Across every session, roll-axis orientations produce a combined MAE of
165.0~cm, whereas pitch-axis orientations produce a MAE of 140.9~cm,
a ratio of 1.24. This is supported by the BRD2606A's dual printed
antennas being optimized for the horizontal plane. Rolling rotates both
antennas away from their intended polarization alignment with the
reflector, whereas pitching primarily changes elevation angle with less
impact on polarization. The direction of pitch also matters, with
pitching toward the reflector consistently outperforming pitching away
at short range.

\begin{figure}[h!]
\centering
\includegraphics[width=\columnwidth]{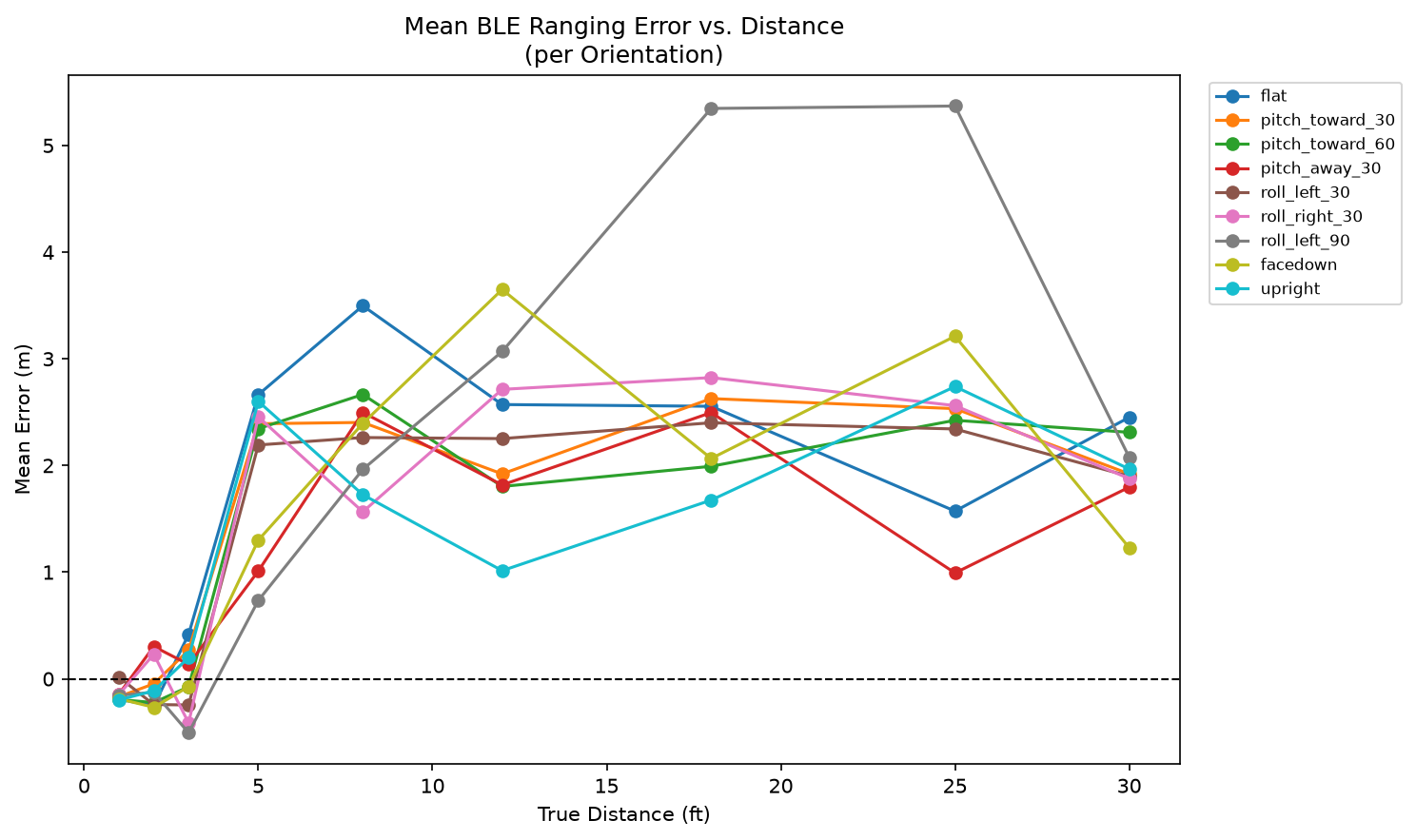}
\caption{Mean ranging error vs. distance per orientation. Roll left 90
diverges sharply at longer distances while other orientations plateau.}
\label{fig:distscale}
\end{figure}

Fig.~\ref{fig:distscale} shows mean error vs. distance for each
orientation. Rolling left 90 degrees continues to show degrading
accuracy at larger distances, while most orientations plateau after
12 to 18~ft. This suggests that rolling left 90 degrees introduces a
severe antenna-alignment factor beyond the general multipath
environment.

\FloatBarrier

\subsection{IMU Tilt Correlation}

This study also observed a linear relationship between ranging error
and tilt angle derived from the IMU measurements. The regression model
produced $r=0.126$ for signed error and $r=0.156$ for absolute error,
both with $p<0.001$ and thus statistically significant at
$N=44{,}576$. This shows that IMU-derived orientation features contain
statistically significant information about CS ranging error, which
prompted the correction model described in Section~IV-E. The gyroscope
magnitude shows a negative correlation with ranging error
($r=-0.136$, $p<0.001$), but this relationship is not meaningful:
boards propped at low-error orientations using physical objects
naturally introduce slight vibrations, which increases gyroscope
readings despite the ranging quality being good.

\begin{figure}[h!]
\centering
\includegraphics[width=\columnwidth]{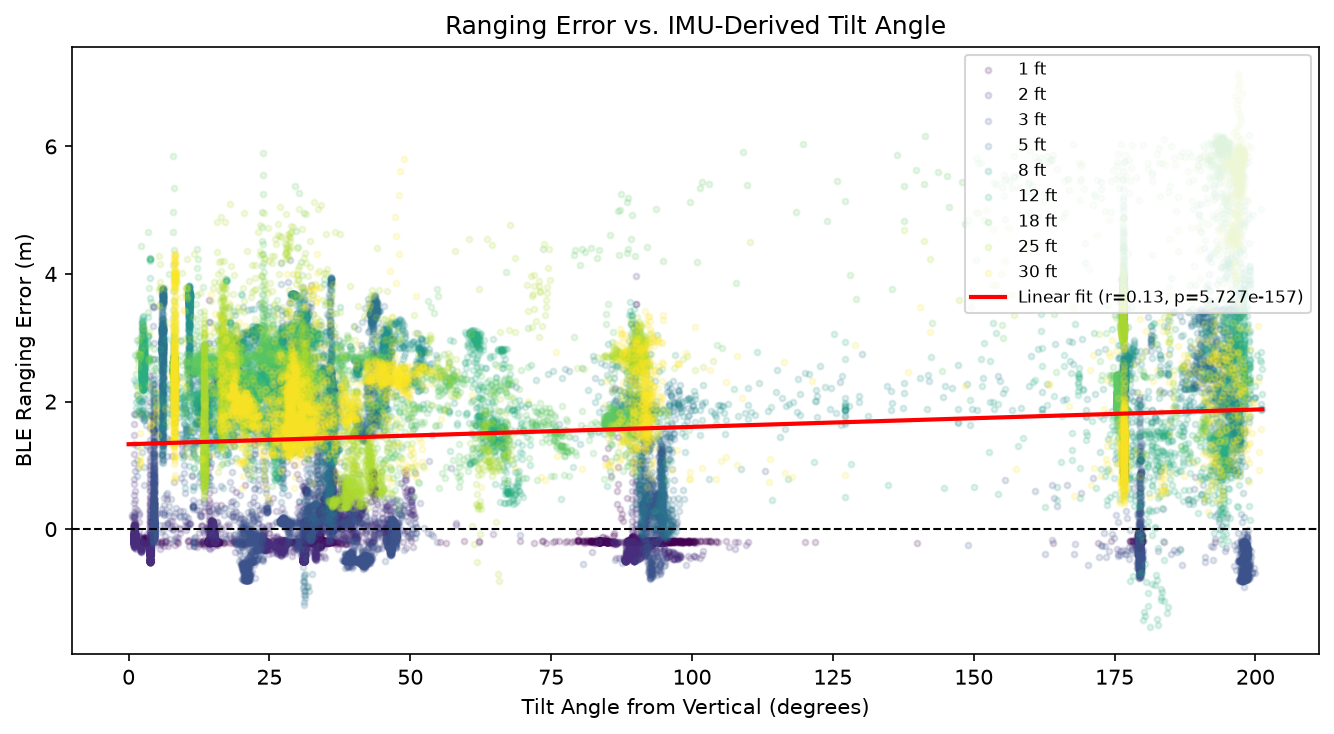}
\caption{Ranging error vs. IMU-derived tilt angle, colored by distance.
Linear fit: $r=0.126$ signed, $r=0.156$ absolute ($p<0.001$).}
\label{fig:tilt}
\end{figure}

Fig.~\ref{fig:tilt} shows ranging error plotted against IMU-derived
tilt angle. The positive slope of the linear fit shows that larger
tilt angles are associated with larger absolute errors.

\FloatBarrier

\subsection{IMU-Based Correction}

As explained in Section~IV-D, the IMU features carry a statistically
significant correlation with ranging error, which prompted the use of
Machine Learning to correct that error using IMU measurements. A random
80/20 split gives 96.9\% MAE reduction for the Random Forest, but
this result is unreliable because consecutive measurements at the same
orientation and distance are almost identical, making it very likely
that similar data points appear in both training and testing. We
therefore report two cross-validation strategies to obtain an honest
measure of generalization.

To select the best corrective approach, we compared three approaches
under LOOO cross-validation. As shown in Table~\ref{tab:models},
Random Forest substantially outperformed both Linear Regression and a
more complex Multilayer Perceptron, achieving a 74.6\% MAE reduction
without having an inflated accuracy.

\begin{table}[h!]
\caption{Model Comparison under LOOO Cross-Validation}
\label{tab:models}
\centering
\small
\setlength{\tabcolsep}{6pt}
\begin{tabular}{lr}
\toprule
Model & LOOO MAE Reduction \\
\midrule
Linear Regression   & 56.6\% \\
MLP (64$\times$64)  & 30.7\% \\
Random Forest       & \textbf{74.6\%} \\
\bottomrule
\end{tabular}
\end{table}

\FloatBarrier

A Leave-One-Orientation-Out (LOOO) cross-validation trains on all
orientations except one, tests on the completely held-out orientation,
and averages across all nine tests. The Random Forest achieves 74.6\%
MAE reduction across all nine orientations, with individual orientation
improvements ranging from 61.5\% to 83.7\%. This is especially
significant since the model is evaluated on completely unseen
orientations: despite only training on eight orientations, it is able
to correct ranging errors for an orientation it has never seen. This is
especially useful in applications such as keyless entry, where the
object will undoubtedly be rotated in several ways.

\begin{figure}[h!]
\centering
\includegraphics[width=\columnwidth]{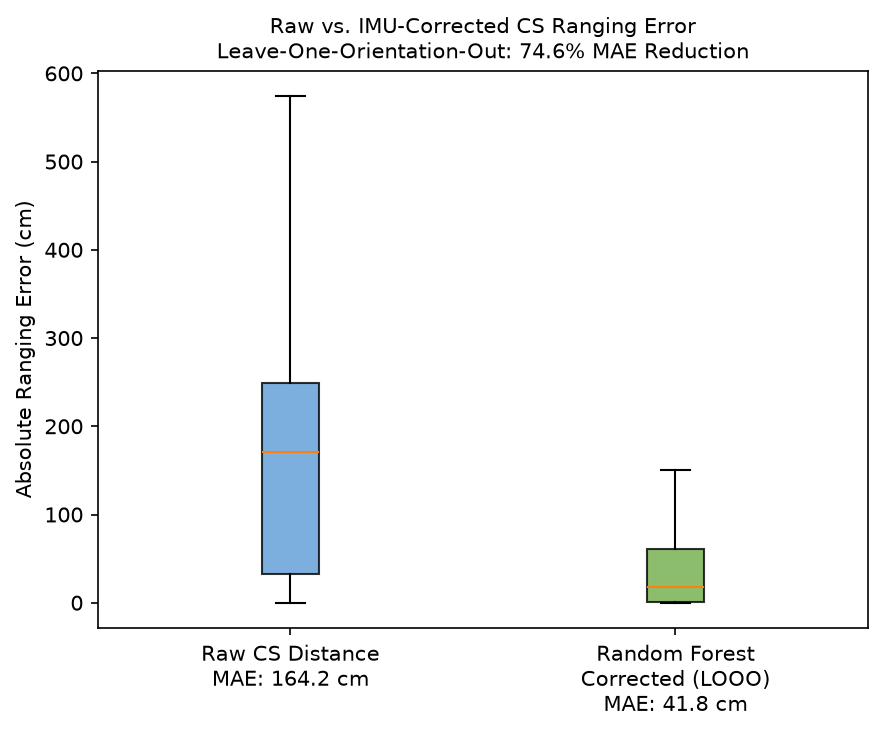}
\caption{Raw CS ranging error vs. corrected error under LOOO
cross-validation. The Random Forest achieves 74.6\% MAE reduction,
generalizing to all orientation categories unseen during training.}
\label{fig:corrected}
\end{figure}

A Leave-One-Distance-Out (LODO) evaluation trains on all distances
except one and tests on the held-out distance, averaging across all
nine distances. The Random Forest achieves an average 16.1\%
MAE reduction across all distances. Fig.~\ref{fig:corrected} shows the
error distributions before and after correction.

\section{Discussion}

Two main device components explain the effects orientation has on
accuracy. First, the BRD2606A's antennas are designed to radiate the
strongest signal when the board is flat. When the board is rolled or
tilted, it points the strongest signal direction away from the
reflector, which lowers the SNR and weakens the direct signal. Thus,
the multipath signals make up a larger share of the total signal the
reflector board receives~\cite{r7,r24}, increasing the ranging error.

Our 74.6\% LOOO reduction shows that the Random Forest generalizes well
to unseen orientations, and the 16.1\% LODO reduction shows that the
model also applies to unseen distances. The smaller LODO gain is
somewhat expected, since distance affects raw ranging error more
strongly than rotation does.

\section{Conclusion}

In this study, we present an evaluation measuring how device orientation
can affect the accuracy of BLE Channel Sounding measurements. Using IMU
measurements across nine orientations at nine distances, our results
show that orientation has a significant impact on CS ranging
performance. We also show that the already present onboard IMU on
commercial hardware has enough information to meaningfully correct this
error using Machine Learning approaches.

The applications of this study are important for any utilization of CS,
in which the devices are handled in unrestricted orientations, such as
keyless vehicle entry, asset tracking, and location based access control
systems. Because of the little to none additional cost for IMU
measurements, deployment is straightforward for this application. Future
work should explore data collection across more environments to improve
generalization and investigate the effect of reflector orientation on CS
ranging accuracy, which was kept fixed in this study.

\bibliographystyle{IEEEtran}

\end{document}